\documentclass[twocolumn,twocolappendix]{aastex631}
\usepackage{amsbsy}
\usepackage{amsfonts}
\usepackage{amssymb}
\usepackage{amsmath}
\usepackage{graphicx}

\begin{document}

\title{Experimental study of Alfv\'en wave reflection from an Alfv\'en-speed gradient relevant to the solar coronal holes}

\author[0000-0001-8093-9322]{Sayak Bose}
\affiliation{ Princeton Plasma Physics Laboratory,  100 Stellarator Road, Princeton, NJ 08540 USA}
\affiliation{Department of Astrophysical Sciences, Princeton University, NJ 08544 USA}
\email{sbose@princeton.edu}

\author[0000-0003-0143-951X]{Jason M. TenBarge}
\affiliation{Department of Astrophysical Sciences, Princeton University, NJ 08544 USA}

\author[0000-0002-5741-0495]{Troy Carter}
\affiliation{Department of Physics and Astronomy, University of California, Los Angeles, California 90095, USA}

\author[0000-0001-7748-4179]{Michael Hahn}
\affiliation{Columbia Astrophysics Laboratory, Columbia University, 550 West 120th Street, New York, NY 10027 USA}

\author[0000-0001-9600-9963]{Hantao Ji}
\affiliation{ Princeton Plasma Physics Laboratory,  100 Stellarator Road, Princeton, NJ 08540 USA}
\affiliation{Department of Astrophysical Sciences, Princeton University, NJ 08544 USA}

\author[0000-0001-6835-273X]{James Juno}
\affiliation{ Princeton Plasma Physics Laboratory,  100 Stellarator Road, Princeton, NJ 08540 USA}

\author[0000-0002-1111-6610]{Daniel Wolf Savin}
\affiliation{Columbia Astrophysics Laboratory, Columbia University, 550 West 120th Street, New York, NY 10027 USA}

\author[0000-0002-6500-2272]{Shreekrishna Tripathi}
\affiliation{Department of Physics and Astronomy, University of California, Los Angeles, California 90095, USA}

\author[0000-0002-6468-5710]{Stephen Vincena}
\affiliation{Department of Physics and Astronomy, University of California, Los Angeles, California 90095, USA}



\begin{abstract}

 We report the first experimental detection of a reflected Alfv\'en wave from an Alfv\'en-speed gradient under conditions similar to those in coronal holes. The experiments were conducted in the Large Plasma Device at the University of California, Los Angeles. We present the experimentally measured dependence of the coefficient of reflection versus the wave inhomogeneity parameter, i.e., the ratio of the wavelength of the incident wave to the length scale of the gradient.  Two-fluid simulations using the {\tt Gkeyll} code qualitatively agree with and support the experimental findings. Our experimental results support models of wave heating  that rely on wave reflection at low heights from a smooth Alfv\'en-speed gradient to drive turbulence. 

\end{abstract}



\section{Introduction} \label{sec:intro}

Coronal holes are low density regions of the solar atmosphere with open magnetic field lines that extend into interplanetary space. These regions appear as dark areas resembling holes in ultraviolet and X-ray images. Spectroscopic measurements indicate that coronal holes are at $10^{6}~\rm{K}$, approximately 200 times hotter than the underlying photosphere \citep{fludra1999electron}. Despite several decades of research, the physics behind the heating of coronal holes is not well understood \citep{Cranmer2009, Cranmer:2015}.

Alfv\'en waves are posited to play a significant role in  the heating of coronal holes \citep{alfven1942existence,mcintosh2011alfvenic}. Perturbations of the foot points of open magnetic field lines, due to the sloshing of the photospheric plasma,  are thought to excite the Alfv\'en waves \citep{priest_2014}.  These waves predominantly travel along the magnetic field lines, transporting  from the photosphere to the corona the energy necessary for heating coronal holes \citep{Cranmer2009}.

Recent observations have found Alfv\'enic waves at the base of coronal holes that satisfy the energy budget needed to heat the plasma in coronal holes \citep{mcintosh2011alfvenic}. The term ``Alfv\'enic'' highlights that in addition to  Alfv\'en waves, transverse kink modes may also be present in coronal holes  \citep{van2008detection, goossens2009nature, goossens2012surface}. Furthermore, the detection of strong damping of Alfv\'enic waves at low heights in coronal holes suggests that wave-driven processes may be responsible for heating the plasma \citep{bemporad2012spectroscopic, hahn2012evidence,hahn2013observational,hara2019nonthermal}.

The most promising wave-based model put forward to explain the damping of wave energy in coronal holes involves nonlinear interaction between counter-propagating waves,  resulting in the development of turbulence, which leads to an irreversible cascade of wave energy to smaller scales, at which the waves are more easily damped, leading to the heating of the plasma \citep{moore1991alfven,Moore_1991,Matthaues1999,Dimtruk_2001,oughton_2001,cranmer2007self,chandran2009alfven}. Recent observations of counter-propagating Alfv\'{e}nic waves in coronal holes support the wave-turbulence model \citep{morton2015investigating}. But, the mechanism responsible for the generation of counter-propagating waves is still unknown. Among the different theories put forward to explain the counter-propagating waves \citep{musielak1992klein, del2001parametric}, a number of them invoke partial reflection of outward propagating Alfv\'{e}n waves at low heights in coronal holes due to an Alfv\'en-speed inhomogeneity along the ambient magnetic field lines \citep{moore1991alfven,1992ApJMoore,musielak1992klein,hahn2018density,asgari2021effects,hahn2022evidence}.

Several experiments have been carried out in the past where Alfv\'en waves were made to propagate through an Alfv\'en-speed gradient produced by a magnetic field gradient. However, none of those experiments detected a reflected wave \citep{stix1958experiments, swanson1972rf, breun1987stabilization, yasaka1988icrf, roberts1989m, vincena2001shear, mitchell2002laboratory, bose2019measured}. An experiment on Alfv\'en wave propagation through multiple magnetic field wells reported a possible indirect signature of wave reflection \citep{zhang2008spectral}, but the  geometry studied is not relevant to coronal holes. 

Here, we report the first direct detection of a reflected Alfv\'en wave in the laboratory under conditions relevant to coronal holes.  The wave experiments were performed in the Large Plasma Device (LAPD; \citet{gekelman2016upgraded, bose2019measured}).    

The rest of this paper is organized as follows.  A comparison of coronal hole and LAPD parameters relevant to Alfv\'en wave physics is described in Section~\ref{appen_A}. The  wave experiments and simulations are presented and analyzed in Sections~\ref{sec:exp} and \ref{sec:two_fluid}, respectively.  In Section~\ref{sec:discussion}, we discuss the implication of our results in the solar context. This is followed by a summary in Section~\ref{sec:summary}.

\section{Alfv\'en waves in coronal holes and LAPD }\label{appen_A}

\begin{deluxetable}{ccCrlc}
	\tablecaption{Dimensionless parameters for coronal holes and LAPD}
	\tablecolumns{6}
	\tablenum{1}
	\tablewidth{0pt}
	\tablehead{
		\colhead{Parameter} &
		\colhead{Coronal hole} &
		\colhead{LAPD}   
	}
	\startdata
	$\bar{\beta}$ & $3 - 18 $  & 5 - 11   \\
	$\bar{\omega}$ & $ \lesssim 9.4 \times 10^{-5} $ & 0.2 - 0.35 \\
	$k_{\perp}^{2} \rho_{\rm i}^{2}$ & $\ll 1^{a}$ & \ll 1   \\
	$k_{\perp}^{2} \rho_{\rm s}^{2}$ & $\ll 1^{a}$ & \ll 1   \\
	$\beta_{\rm e}$ & $ 1.5 - 9.6 \times 10^{-3}$ & 0.7 - 1.5 \times 10^{-3}    \\
	${\lambda_{\parallel}}/{L_{\rm A, min}}$ & $\gtrsim 4.5 $  &  \lesssim 7.5   \\
	$\nu_{\rm ei}/\omega$ & $ 2 - 2400 $  &  42 - 114   \\
	$L_{\rm A, min}/\lambda_{\rm mfp, e}$ & $\sim 13 $  &  9 - 46   \\ 
	$b/B_{0}$ & $\lesssim 0.02$  &  \lesssim  10^{-4}   \\
	\enddata
	\tablenotetext{a}{Assuming that Alfv{\'e}n waves in coronal holes satisfy nearly ideal MHD conditions.}
	\label{table:lab_corona}
\end{deluxetable}

The plasma and Alfv\'en wave parameters in LAPD were scaled to match those in coronal holes, to within laboratory limitations, by adpoting the frame work discussed by \citet{bose2019measured}. In the Sun, Alfv{\'e}n waves excited in the photosphere \citep{narain1996chromospheric, priest_2014} propagate upward through coronal holes along the ambient magnetic field lines, which are nearly straight. The geometry of LAPD is similar to coronal holes. LAPD is a cylindrical machine, wherein we excite waves at one end of the machine, and the waves follow the magnetic field along the length of the machine.

Alfv\'en waves interact with electrons and ions in the plasma as the waves propagates. The response of the electrons to the wave field is characterized by the dimensionless parameter, $\bar{\beta} = 2v_{\rm{te}}^{2}/v_{\rm{A}}^{2}$. The electron thermal speed $v_{\rm{te}}=\sqrt{T_{\rm{e}}/m_{\rm{e}}}$, where $T_{\rm{e}}$ is the electron temperature in energy units and $m_{\rm{e}}$ is the electron mass. The Alfv\'en speed   $v_{\rm{A}}=B_{\rm 0}/\sqrt{\mu_{0}\rho}$,  where  $B_{0}$ is the ambient magnetic field, $\mu_{0}$ is the permeability of free space, $\rho=\left( n_{\rm i} m_{\rm i} + n_{\rm e} m_{\rm e} \right)$ is the mass density of the plasma, $n_{\rm i}$ is the ion number density, $n_{\rm e}$ is the electron number density, and $m_{\rm i}$ is the ion mass.

In coronal holes, $\bar{\beta} > 1$. As a result, the electrons respond adiabatically to the wave field.  We matched the condition of coronal holes  $\bar{\beta}> 1$  in LAPD by tuning the parameters such as $n$, $T_{\rm e}$ and $B_{0}$ (See Table~\ref{table:lab_corona}).

 The wave energy in coronal holes has a wide spectrum. However, most of the wave energy occurs at  $\omega \ll \omega_{\rm ci}$, where  $\omega_{\rm{ci}}=qB_{0}/m_{i}$ is the angular ion cyclotron frequency, and $q$ is the charge of the ion.  Models suggest photospheric fluctuations primarily generate waves at frequencies $f=\omega/2\pi$  between  $\sim 0.1 - 100~\rm{mHz}$ \citep{cranmer2005generation}. The  ambient magnetic field in a coronal hole is $\sim 0.7\;\rm G$ at a height of $0.15\;\rm R_{\odot}$ \citep{morton2015investigating}, where $\rm R_{\odot}$ is the solar radius. At this height, $\bar{\omega} = \omega / \omega_{ci}$ ranges from \mbox{ $\approx 9.4\times 10^{-8} - 9.4 \times 10^{-5}$}.   The parameter $\bar{\omega}$ reduces the Alfv\'en wave speed as $\omega$ approaches $\omega_{\rm{ci}}$. This effect can be seen from the  simplified Alfv\'en wave dispersion relation adopted from  \citet{gekelman1997laboratory, gekelman2011many},
\noindent
\begin{equation}
    \frac{\omega}{k_{\parallel}}\approx v_{\rm A}\sqrt{1-\bar{\omega}^{2}},
    \label{KAW_dispersion2}
\end{equation}
\noindent
where $\rho_{\rm s}=c_{\rm s}/\omega_{\rm ci}$ is  the ion sound gyroradius, $c_{\rm s}=\sqrt{T_{\rm e}/ m_{\rm i}}$ is the ion sound speed, $k_{\parallel}=2\pi/\lambda_{\parallel}$ is the wave number parallel to $\mathbf{B_{0}}$, and $\lambda_{\parallel}$ is the wavelength along $\mathbf{B_{0}}$. We minimized the finite frequency correction, $1-\bar{\omega}^2$, by limiting the wave frequency to satisfy $\bar{\omega}\leq 0.35$ near the antenna.

Alfv\'en wave dynamics can be affected by two-fluid and kinetic effects \citep{cross1988introduction,gekelman1997laboratory,cramer2011physics}. One of the parameters that affects two-fluid and kinetic physics is $k_{\perp}$, which appears in the rigorously derived Alfv\'en wave dispersion relation through the dimensionless terms $k_{\perp}^2 \rho_{\rm i}^2$, $k_{\perp}^{2}\rho_{\rm s}^2$, and $k_{\perp}^2\delta_{\rm{e}}^2$ \citep{gekelman1997laboratory,cramer2011physics,bose2019measured}. Here, $k_{\perp}=2\pi/\lambda_{\perp}$ is the wave number perpendicular to $\mathbf{B_{0}}$, $\lambda_{\perp}$ is the perpendicular wavelength of Alfv\'en wave, $\rho_{\rm{i}}=v_{\rm{ti}}/\omega_{\rm{ci}}$ is the ion Larmor radius, $v_{\rm{ti}}=\sqrt{T_{i}/m_{\rm i}}$ is the ion thermal velocity, $T_{\rm{i}}$ is the ion temperature, $\rho_{\rm{s}}=c_{\rm{s}}/\omega_{\rm{ci}}$ is the ion sound gyroradius,  $\delta_{e}=c/\omega_{\rm{pe}}$ is the electron skin depth, $c$ is the speed of light,   $\omega_{\rm pe}=\sqrt{ne^2/m_{\rm e}\epsilon_{0}}$ is the electron plasma frequency, $e$ is the fundamental unit of electrical charge and $\epsilon_{0}$ is the permitivity of free space.

In coronal holes, there is no measurement of $k_{\perp}$ of the Alfv\'en waves, but most of the wave-based heating models invoke ideal MHD (magnetohydrodynamic) conditions, i.e.,  $k_{\perp}^2 \rho_{\rm i}^2$, $k_{\perp}^{2}\rho_{\rm s}^2 $, and $k_{\perp}^2\delta_{\rm{e}}^2 $ are all $\ll 1$ \citep{gekelman1997laboratory}.  We used an antenna that excites a large dominant $\lambda_{\perp} $ to match this condition in LAPD \citep{gigliotti2009generation,karavaev2011generation}.  In our experiments, the dominant $\lambda_{\perp}$ of the incident wave is approximately 21.78~cm ensuring that  $k_{\perp}^{2}\rho_{\rm i}^2$, $k_{\perp}^{2}\rho_{\rm s}^2$, $k_{\perp}^2\delta_{\rm{e}}^2$ are all $ \ll 1$. This value of $\lambda_{\perp}$ is determined from the wave data using a Fourier-Bessel analysis \citep{churchill1987introduction} as illustrated by \cite{vincena1999}.

The energy in coronal holes is in the magnetic field. The magnetic pressure dominates over thermal pressure, which is represented by the dimensionless parameter $\beta_{\rm e} = 2\mu_0 n T_{\rm e} /B_{0}^2$. The value of $\beta_{\rm e}$ varies between $\approx 9.6\times 10^{-3}$ and $1.5\times 10^{-3}$ at low heights in coronal holes \citep{bose2019measured}. To match this in LAPD, we adjusted $B_{0}$, $n$, and $T_{\rm e}$ to produce a  value of $\beta_{\rm e}$ ranging from $\approx 0.7 \times 10^{-3}$  to $\approx 1.5 \times 10^{-3}$.  

Alfv\'en waves encounter a strong $v_{\rm{A}}$ gradient at low heights in coronal holes. An estimate of the effect of the gradient on a wave is  $\lambda_{\parallel}/L_{\rm{A}}$ which is also referred to as wave inhomogeneity parameter.  $L_{\rm{A}}$ is the length scale of $v_{\rm{A}}$ in the gradient. This length scale is defined as  $v_{\rm A}/v^{\prime}_{\rm A}$, where $v^{\prime}_{\rm A}$ is the first spatial derivative of $v_{\rm A}$. If $\lambda_{\parallel}/L_{\rm{A}} \gtrsim 1$, the gradient is strong because $v_{\rm{A}}$ changes significantly within a wavelength. While if $\lambda_{\parallel}/L_{\rm{A}} < 1$, the gradient is weak because the variation of $v_{\rm{A}}$ within a wavelength is small. In coronal holes,   $L_{\rm{A}}$ varies spatially at low heights and we have used $L_{\rm{A, min}}$ for scaling purposes, where $L_{\rm{A, min}}$ is the minimum value of the  length scale in the gradient region. Note that the length scale attains its minimum value at the strongest part of the gradient. The bulk of the wave energy in coronal holes satisfies the condition $\lambda_{\parallel}/L_{\rm{A, min}} \gtrsim 4.5$ \citep{morton2015investigating, bose2019measured}. To match this regime in LAPD, we adjusted $B_{0}$, $n$, and $\omega$ to produce $\lambda_{\parallel}/L_{\rm{A, min}}$ up to 7.5.

Alfv{\'e}n waves are known to damp due to Coulomb collisions \citep{cramer2011physics}.  A parameter characterizing the effect of collisions on waves is $\nu_{\rm{ei}}/\omega$, where $\nu_{\rm{ei}}$ is the electron-ion collision frequency. The ratio $\nu_{\rm{ei}}/\omega$ is a measure of the number of collisions occuring in a wave period. If $\nu_{\rm{ei}}/\omega \gg 1$, the plasma is collisional for the wave; and if $\nu_{ei}/\omega \ll 1$, it is collisionless plasma. To calculate $\nu_{\rm{ei}}$,  we use the conventional method from \cite{Braginskii:1965}, 
\begin{equation}
    \nu_{\rm{ei}}=2.9 \times 10^{-6} n \Lambda T_{\rm{e}}^{-3/2},
\end{equation}
where $\Lambda$ is the Coulomb logarithm \citep{huba2016nrl}, and $\nu_{\rm{ei}}$ is in Hz. For coronal hole conditions of $n\sim 2\times10^7~\rm{cm^{-3}}$ and $T_{\rm{e}}\sim 86~\rm{eV}$ $\left( 10^6~\rm{K} \right)$ at 0.15 $R_{\odot}$, we find $\nu_{\rm{ei}}\sim 1.5~\rm{Hz}$ \citep{morton2015investigating,Cranmer2009}. Therefore, for wave frequencies ranging from  $f\sim 0.1 - 100~\rm{mHz}$ \citep{cranmer2005generation},  $\nu_{\rm{ei}}/\omega$ varies from $\sim$ 2 to $2400$. Hence, at low heights  in the $v_{\rm{A}}$ gradient region of coronal holes, collisions are sufficient to affect Alfv\'en wave physics. To match this in the laboratory, we adjusted, $\omega$, $n$ and $T_{e}$ (see Table~\ref{table:lab_corona}).

Another estimate of the effect of electron-ion collisions on the Alfv\'en wave in the gradient region is given by the ratio of the electron mean free path, $\lambda_{\rm mfp, e}$, to  $L_{\rm A, min}$ \citep{bose2019measured}. This ratio gives a measure of the number of electron mean free paths within the $v_{\rm{A}}$ gradient.  The value of $L_{\rm{A, min}}/\lambda_{\rm mfp, e}$ in coronal holes is estimated to be $\sim 13$ \citep{bose2019measured}.   We match this in LAPD by suitably selecting the range of $n$, $T_{\rm e}$ and $L_{\rm A, min}$. See Table~\ref{table:lab_corona} for representative values of $L_{\rm{A, min}}/\lambda_{\rm mfp, e}$ in LAPD.

Lastly, Alfv{\'e}nic waves with amplitude as high as $b/B_{0} \sim 0.02$ were reported in coronal holes \citet{mcintosh2011alfvenic}. In this regime, nonlinear effects may play a role. Following \citet{bose2019measured}, we  restricted our experiment to the low amplitude regime $b/B_{0} \lesssim 10^{-4}$ to avoid known nonlinear effects associated with larger amplitude Alfv{\'e}n waves.

\section{Experiments}\label{sec:exp}

\subsection{Overview}

\begin{figure}
    \centering
    \includegraphics[scale=0.85]{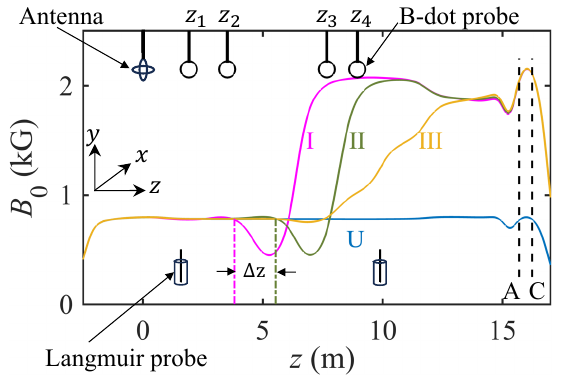}
    \caption{Schematic of the experimental arrangement. The anode (A) and the cathode (C) of the plasma source are at 15.7 and 16.23~m, respectively.   The orthogonal ring antenna used to excite Alfv\'en waves is located at $x=y=z=0$. The vertical ring of the antenna lies in the $yz$-plane, while the horizontal ring lies in the $xz$-plane. The four magnetic field profiles used for the wave experiment (U, I, II, and III) are indicated by different colors. The location of the four B-dot probes are indicated at the top of the figure. A Langmuir probe was used to make measurements at two locations shown at the bottom of the figure. The quantity $\Delta z$ corresponds to 1.72~m.  }
    \label{fig:exp_set_up}
\end{figure}

 LAPD was used to produce an $\approx 19~\rm{m}$ long cylindrical plasma column \citep{gekelman2016upgraded}. Electromagnets arranged coaxially along the cylindrical vacuum vessel were used to generate a background axial magnetic field, $B_{0}$.  Plasma was formed through electron impact ionization of Helium by applying a discharge voltage between a mesh anode and a $\rm{LaB}_{6}$ cathode located at the far end of the machine, as shown in Fig.~\ref{fig:exp_set_up}.

For our experiments, a two cycle linearly polarized Alfv\'en wave of azimuthal mode number $m=1$  was excited using an orthogonal ring antenna placed on the cylindrical axis of LAPD at $x=y=z=0$ \citep{gigliotti2009generation, karavaev2011generation, bose2019measured}. Triaxial B-dot probes at $z_{1}=1.92~\rm{m}$, $z_{2}=3.51~\rm{m}$, $z_{3}=7.67~\rm{m}$, and $z_{4}=8.95~\rm{m}$ were used to  make wave magnetic-field measurements \citep{everson2009design, bose2018understanding}.

We studied reflection of Alfv\'en waves by varying $\lambda_{\parallel}$ and $L_{\rm{A, min}}$. The wave frequency was changed to vary $\lambda_{\parallel}$, while different $B_{0}$ gradients were employed to alter $L_{\rm{A, min}}$. The four $B_{0}$ profiles used for the reflection experiments are shown in Fig.~\ref{fig:exp_set_up}.

A Langmuir probe was used to measure  $n$  and  $T_{e}$ for the gradient cases \citep{hershkowitz1989langmuir, bose2017langmuir}. The density measurements were calibrated using an interferometer that measured line-integrated density. For the gradient cases, the estimated plasma parameters  at $z=1.6~\rm{m}$  are  $n\sim 5-8  \times 10^{12}~\rm{cm^{-3}}$ and $T_{e} \sim 3 ~\rm{eV}$, while  at $z=9.9~\rm{m}$, they are $n\sim 7-9  \times 10^{12}~\rm{cm^{-3}}$ and $T_{e} \sim 10 ~\rm{eV}$.  We do not have Langmuir probe measurements for the uniform $B_{0}$ case; the estimated $n$ from the Alfv\'en wave dispersion relation is $\sim 8\times 10^{12}\:~\rm{cm^{-3}}$.

\subsection{Detection of an Alfv{\'e}n wave reflected from a $v_{\rm{A}}$ gradient}

We performed a number of wave experiments to detect a reflected Alfv\'en wave from a $v_{\rm{A}}$ gradient and validate the findings. We first studied the wave propagation through a uniform $B_{0}$ as a control case (profile U in Fig.~\ref{fig:exp_set_up}). Second, we excited an Alfv\'en wave in the presence of a strong $v_{\rm{A}}$ gradient (profile I in Fig.~\ref{fig:exp_set_up}) and detected a reflected wave from the $v_{\rm{A}}$ gradient. Third, we pushed the $v_{\rm{A}}$ gradient further from the antenna (profile II in Fig.~\ref{fig:exp_set_up}) and found the phase difference between the incident and reflected wave increased, confirming that the $v_{\rm{A}}$ gradient is the reflector. Fourth, we weakened the $v_{\rm{A}}$ gradient significantly (profile III in Fig.~\ref{fig:exp_set_up}); we could not find a reflected wave from this gradient, proving that reflection is a gradient-driven effect. In the subsubsections below, we describe the results and analysis of these experiments using an $\approx 76~\rm{kHz}$ incident wave as an example.

\subsubsection{Alfv\'en wave in uniform $B_{0}$}

\begin{figure}
    \centering
    \includegraphics[scale=0.75]{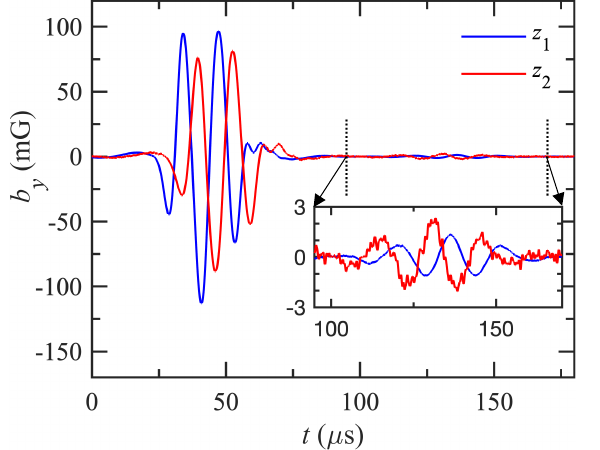}
    \caption{The time variation of the  $y$ component of the wave magnetic field, $b_{y}$.  The measurements were made on the axis of LAPD at $z_1$ and $z_2$ for case U. Initially, the wave at $z_1$ (blue) occurs before the wave at $z_2$ (red), suggesting that the wave is propagating away from the antenna. The waveform within the dotted lines, highlighted in the inset figure, is due to reflection from the cathode and propagating towards the antenna.  }
    \label{fig:uniform}
\end{figure}

Figure~\ref{fig:uniform} shows the wave magnetic field time series recorded by B-dot probes  at $z_{1}$ and $z_{2}$ for gradient U. Initially,   the peaks and troughs of the wave signal at $z_{1}$ leads $z_{2}$, which is consistent with an incident wave propagating in the $+z$ direction. Some time after the incident wave in Fig.~\ref{fig:uniform}, we see a much smaller wave in the region between the dotted lines. The signal of this smaller wave is observed at $z_{2}$ before $z_{1}$, indicating the waveform is propagating in the $-z$ direction, towards the antenna.  

\begin{figure*}
    \centering
    \includegraphics[scale=0.75]{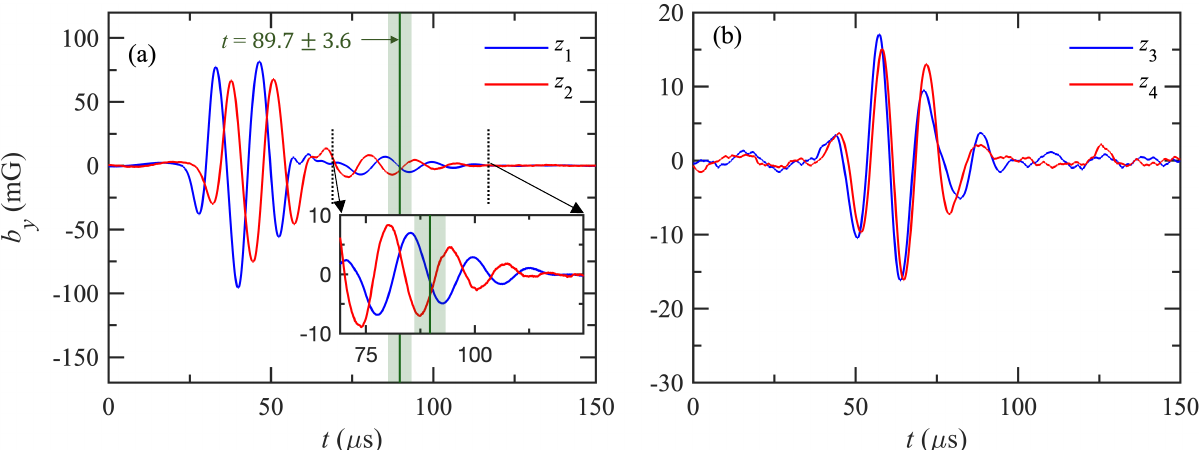}
    \caption{The $b_y$ time series for gradient I as measured on the  axis of LAPD (a) on the low field side at $z_1$ and $z_2$, and (b) on the high field side at $z_{3}$ and $z_{4}$. The inset figure in (a) shows the waveform propagating towards the antenna on the low field side. The green line and shaded region in (a) represent the expected time for a wave refelcted from the cathode to arrive at $z_2$. }
    \label{fig:reflection_grad1}
\end{figure*}

Our analysis suggests that the smaller wave signal is reflected from the plasma source at the far end of LAPD. Furthermore, reflection introduces an additional $180^{\circ}$ phase difference between the reflected and incident wave; the smaller signal in Fig.~\ref{fig:uniform} has three peaks and two troughs, while in the incident wave has two peaks and three troughs. We performed time-of-flight  calculations using the logic that if the smaller wave at $z_{1}$ is reflected from the plasma source, then the time lag  between the incident and the inverted smaller wave should be equal to twice the time of flight of the incident wave from $z_{1}$ to the plasma source. The time lag ($t_{\rm{lag}}^{\rm{U}}$) between the incident and the  smaller waves obtained from their phase difference at $z_{1}$ is $95.3\pm0.4\:~\rm{\mu s}$. The error bar represent the $1\sigma$ uncertainty obtained by calculating the standard deviation of the time lag measured at various radial locations. The time of flight ($t_{\rm{tof}}^{\rm{U}}$) of the incident wave from $z_{1}$ to the plasma source determined by dividing the distance between $z_{1}$ and the cathode, 14.31~m, by the velocity of the wave. The velocity of the wave is $3\pm 0.1\times10^{5}~\rm{m~s^{-1}}$, as determined by dividing the distance between $z_{1}$ and $z_{2}$ by the time lag between the wave signals at those two locations.  Therefore, $t_{\rm{tof}}^{\rm{U}}$ is $47.7\pm1.6~\mu \rm{s}$. Hence, $t_{\rm{lag}}^{\rm{U}} $ agrees with $2t_{\rm{tof}}^{\rm{U}}$,  indicating that the smaller wave has been reflected from the plasma source. Our findings agree with previous experiments where physical objects in the plasma, such as metal and insulator plates, were found to reflect Alfv\'en waves \citep{leneman2007reflection}.

\subsubsection{Wave reflection from a strong $v_{\rm{A}}$ gradient}\label{subsec:strong_va}

We introduced a strong $v_{\rm{A}}$ gradient between the antenna and the far end of LAPD and detected a wave  reflected from the $v_{\rm{A}}$ gradient.  The minimum value of the $v_{\rm{A}}$  length scale within the gradient was $L_{\rm{A, min}}= 0.74~\rm{m}$. The wave magnetic field data before the gradient is given in Fig.~\ref{fig:reflection_grad1}(a). Initially, the peaks and troughs of the  wave signal at $z_{1}$  leads $z_{2}$, which is expected for an incident wave. However, in the region between the dotted lines, a small wave signal is observed at $z_{2}$ before $z_{1}$, suggesting the wave is propagating in the $-z$ direction. This is likely due to a partial superposition of waves reflected from gradient I and the plasma source, as we explain shortly. Here we remind the reader that since $v_{\rm A} \propto B_{0}$, the time window for the wave reflected by the plasma source at $z_{1}$ and $z_{2}$ for the strong $v_{\rm A}$ gradient and uniform $B_{0}$ profile will be different as the values of $B_{0}$ near the plasma source is different in the two cases.

In order to help disentangle wave reflection from the gradient from wave reflection by the plasma source, we also measured the wave magnetic field on the high field side at $z_{3}$ and $z_{4}$ (Fig.~\ref{fig:reflection_grad1}(b)). The first two peaks and  troughs at $z_{3}$ leads $z_{4}$, which is consistent with wave propagation in the $+z$ direction. However, the third trough of $z_{3}$ lags $z_{4}$, suggesting the presence of a wave reflected from the far end in the measured signal. The amplitude of the third trough at $z_{4}$ is also more than at $z_{3}$, suggesting constructive interference between waves travelling in the $+z$ (incident) and $-z$ (reflected from far end) directions. The presence of a far-end reflected wave is also supported by the amplitude of  the third peak at $z_{4}$, which is greater than $z_{3}$. This observation suggests that the far-end reflected wave may have reached $z_{4}$ around $t=68.3\pm3.5~\mu \rm{s}$. The upper  limit   corresponds to the time  of the second peak at $z_{4}$, where we observe a signature of a reflected wave from the far end.  The lower limit refers to the time  of the second trough at $z_{4}$ where we do not observe a sign of reflection from the far end.  

Building on this, we put forward two sets of analyses that imply that  the first cycle of the wave propagating in the $-z$ direction at $z_{2}$ is reflected from gradient I only and does not have any contribution due to far-end reflection. First, we estimate the time at which a wave reflected from the plasma source will reach $z_{2}$. The arrival time  of the reflected wave from the source  is equal to $68.3\pm3.5~\mu s$ plus the time of flight of the wave from $z_{4}$ to $z_{2}$.  Since the speed of Alfv\'en wave propagation in the $+z$ and $-z$ direction is the same, the time of flight of an Alfv\'en wave from $z_{4}$ to $z_{2}$ (i.e., the $-z$ direction) is determined from the phase difference of the incident wave (i.e., the $+z$ direction) between $z_{2}$ and $z_{4}$. The inferred time of flight is $21.4\pm0.7~\mu \rm{s}$. Therefore, the part of the  waveform at $z_{2}$ beyond $t=89.7\pm3.6 ~\mu \rm{s}$ may contain a contribution from a wave reflected from the plasma source. Hence, the first cycle of the waveform in the inset graph of Fig~\ref{fig:reflection_grad1}(a) is not affected by reflection from the source. 

Next, we show that the location of reflection of the first cycle of the wave propagating in the $-z$ direction at $z_{2}$ is within gradient I. Since  gradient I is located between  $z_{2}$ and $z_{3}$, for the first cycle to be reflected from the gradient I, the time lag between the incident and reflected wave at $z_{2}$, $t_{\rm lag}^{\rm I}$, must be less than the twice the time of flight of the wave from $z_{2}$ to $z_{3}$, $2t_{\rm tof}^{\rm I}$. We used the results of the simulations given in Section~\ref{sec:two_fluid} to correlate the phases of incident and gradient-reflected wave.  Simulations showed that reflection from a $v_{\rm{A}}$ gradient introduces a $180^{\circ}$ phase difference between the incident and the reflected wave. In addition, the simulations also suggest that the leading low amplitude trough of the incident Alfv\'en wave may not produce a sufficiently strong reflected wave to be experimentally detectable. The subsequent peaks and troughs of the incident Alfv\'en wave, though, are predicted to  produce a detectable reflected wave. Therefore, in Fig.~\ref{fig:reflection_grad1}(a), the first peak of the incident wave should correspond to the first trough of the wave moving in the $-z$ direction if it is reflected from the $v_{\rm{A}}$ gradient. The time lag between the first peak of the incident wave and the first trough of the reflected wave  is $t_{\rm{lag}}^{\rm{I}} = 36.0\pm0.2~\mu \rm{s}$. The time of flight of the wave from $z_{2}$ to $z_{3}$ obtained by comparing the phase difference between the $+z$ propagating part of the wave at $z_{2}$ and $z_{3}$ is $t_{\rm{tof}}^{\rm{I}}=20.4 \pm 0.7~\mu \rm{s}$. Hence, $t_{\rm{lag}}^{\rm{I}} < 2 t_{\rm{tof}}^{\rm{I}}$ indicating that the source of reflection is located within gradient I. This experimental finding is  supported by our simulations described in Section~\ref{sec:two_fluid}.

\subsubsection{Wave reflection experiment by changing the 
location of a strong $v_{\rm{A}}$ gradient}\label{sub_sec:grad_shifted}

\begin{figure}[t]
    \centering
    \includegraphics[scale=0.75]{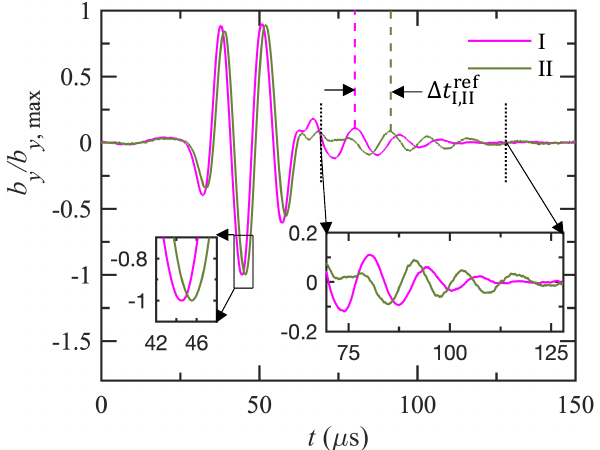}
    \caption{Comparison of the normalized $b_{y}$ time series  measured at $z_{2}$ on the axis of LAPD before (magenta) and after (green) moving the gradient away from the antenna. The inset figure on the right highlights the significant phase difference  observed between the reflected waves due to moving the gradient away from the antenna. The inset figure on left shows that there is a minor phase difference between the incident waves of the two cases. }
    \label{fig:grad_moved}
\end{figure}

To further confirm experimentally that gradient I is indeed reflecting an Alfv\'en wave, we moved  the gradient further away from the antenna and checked if the time lag between the incident and reflected wave increased accordingly. We refer to this shifted gradient as gradient II. A comparison of the wave signal at $z_{2}$ for gradients I and II is presented in Fig.~\ref{fig:grad_moved}.

We performed further analysis to quantitatively show that the phase difference between the reflected wave from gradients I and II  agrees with the extra distance traversed by the wave to reach gradient II and return. The investigation was done using the first  wave cycle propagating in $-z$ direction for both gradients. This is because an analysis for gradient II following the methodology in subsection~\ref{subsec:strong_va} that was used for gradient I showed that the reflector for the first cycle is located within the gradient II region.   The time lag between waves reflected from I and II is $t_{\rm{lag}}^{\rm{I, II}}\approx 2\Delta z/v_{\rm{ph},\parallel}$, where $\Delta z = 1.72~\rm{m}$ is the distance between gradients I and II (Fig.~\ref{fig:exp_set_up}), and $v_{\rm{ph},\parallel}$ is the phase velocity of the Alfv\'en wave parallel to $B_{0}$ in the low-field side. We find $v_{\rm{ph},\parallel}$ = $3.3\pm 0.1 \times 10^{5} ~\rm{m~s^{-1}}$, as determined by dividing the distance between $z_{1}$ and $z_{2}$ by the phase difference of the incident wave at $z_{1}$ and $z_{2}$. This gives, $t_{\rm{lag}}^{\rm{I, II}} \approx 10.4\pm0.3~ \mu \rm{s}$. The phase difference between the reflected wave from I and II is $\Delta t_{\rm{I, II}}^{\rm{ref}}\approx 11.0\pm0.2~ \mu \rm{s}$. In Fig.~\ref{fig:grad_moved}, we find a minor phase offset between the incident waves for cases I and II possibly due to small difference in density. This minor phase offset adds to the phase difference between the reflected waves from the two gradients. Correcting for the phase difference between incident waves, we find $\Delta t_{\rm{I, II}}^{\rm{ref}}\approx 10.0\pm 0.3 ~\mu \rm{s}$. Thus, $t_{\rm{lag}}^{\rm{I,II}}$ closely agrees with $\Delta t_{\rm{I, II}}^{\rm{ref}}$ further confirming that the gradient is reflecting Alfv\'en waves. This experimental finding is supported by our simulation discussed in Section~\ref{sec:two_fluid}.

\subsubsection{Alfv\'en  wave incident on a weak $v_{\rm{A}}$ gradient}\label{subsec:weak_va}

\begin{figure}
    \centering
    \includegraphics[scale=0.75]{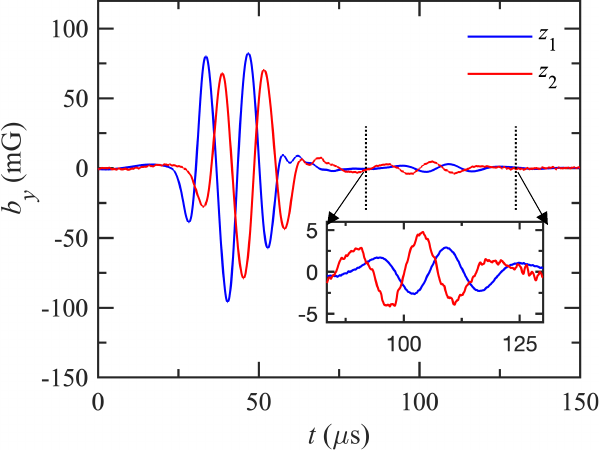}
    \caption{The time variation of  $b_y$ for gradient III measured on the axis of LAPD at $z_1$ and $z_2$. Initially, the peaks and troughs of the wave signal at $z_{1}$ leads $z_{2}$, which is consistent for a wave propagating in the $+z$ direction. However, some time after the incident wave, the smaller wave reflected from the far end is seen propagating in the $-z$ direction.}
    \label{fig:weak_grad}
\end{figure}

As a further test to prove that the first cycle of the reflected wave in Fig.~\ref{fig:reflection_grad1}(a) is a gradient driven effect, we weakened the $v_{\rm{A}}$ gradient and repeated the experiment.  The $L_{\rm{A, min}}$  within the gradient was $ 3.24~\rm{m}$. The wave magnetic field data at $z_{1}$ and $z_{2}$ for III is shown in Fig.~\ref{fig:weak_grad}. Initially, the troughs and peaks of the wave signal at $z_{1}$ leads $z_{2}$, which is consistent for a wave propagating in the $+z$ direction. However, some time after the incident wave in Fig.~\ref{fig:weak_grad}, we see a much smaller wave in the region between the dotted lines. This small wave is propagating in the $-z$ direction, as the peaks and troughs of the small wave at $z_{2}$ leads $z_{1}$.   

Analysis shows that the small wave is reflected from the plasma source of LAPD and that there is no signature of the wave being reflected from the gradient. For the small wave to be reflected from the source, the time lag ($t_{\rm{lag}}^{\rm{III}}$) between the incident and small wave at $z_{1}$  should be equal to twice the time of flight ($t_{\rm{tof}}^{\rm{III}}$) of the wave from the $z_{1}$ to the plasma source. The $t_{\rm{lag}}^{\rm{III}}$ measured by cross-correlating the incident and small wave after accounting for $180^{\circ}$ phase shift caused by reflection is $67.2\pm 1.5~\mu \rm{s}$. The time of flight calculated using $t_{\rm{tof}}^{\rm{III}}=\int_{z=z_{1}}^{z=16.3} dz/\int_{z=z_{1}}^{z=16.3} v_{\rm{A}}\,dz$ is $35.7\pm 1.9\,\mu \rm{s}$. Here, the $z$ dependence of $B_{0}$ primarily causes an axial variation of $v_{\rm{A}}$ as $v_{\rm{A}}\propto B_{0} $, while a minor variation in $n$ doesn't effect $v_{\rm{A}}$ much as $v_{\rm{A}} \propto 1/\sqrt{n}$. Therefore, $t_{\rm{lag}}^{\rm{III}}$ agrees with $2t_{\rm{tof}}^{\rm{III}}$ to within the measurement uncertainty, proving that the small wave is reflected from the plasma source. Since we do not observe any other wave between the incident wave and the small source-reflected wave in Fig.~\ref{fig:weak_grad}, we conclude that gradient III does not reflect Alfv\'en waves. This experimental finding is supported by our simulations, which shows that gradient III does not reflect (see Section~\ref{sec:two_fluid}).

\subsection{Dependence of reflected Alfv\'en wave energy on the $v_{\rm{A}}$ inhomogeneity}

We determined the coefficient of reflection, $\mathcal{R}$, to quantify the effectiveness of a $v_{\rm{A}}$ gradient in reflecting Alfv\'en waves. As a first step for determining $\mathcal{R}$, we calculate the energy of the incident wave and of the wave reflected from the $v_{\rm{A}}$ gradient. The energy,  $\mathcal{E}$, of an Alfv\'en wave is obtained by integrating the Poynting vector of the wave and $\mathcal{E}$ is given by 
\begin{equation}
    \mathcal{E} =\int \left(  \iint S_{\parallel} \, dxdy \right)dt, \label{Eq:wave_energy}
\end{equation}
where $S_{\parallel}\approx \frac{1}{4\pi}b^{2} v_{\rm{ph, \parallel}}$ is the energy flux along $\mathbf{B_{0}}$ \citep{bose2019measured}. The spatial integration is carried out over the $xy$ cross section of LAPD, and the integration in time, $t$, is carried out over the duration of the wave train. 

 $\mathcal{R}$ is obtained from $\mathcal{E}$ using   
\begin{equation}
    \mathcal{R}=\frac{\it{\Gamma}_{\rm{r}}}{\it{\Gamma}_{\rm{i}}}=\frac{\mathcal{E}_{\rm{r}}/t_{\rm{dur,r}}}{\mathcal{E}_{\rm{i}}/t_{\rm{dur,i}}}=\frac{\left[ \int \left(  \iint b_{\rm{r}}^2 \, dxdy \right)dt \right]/t_{\rm{dur,r}}}{\left[ \int \left(  \iint  b_{\rm{i}}^2\, dxdy \right)dt \right] /t_{\rm{dur,i}}},
    \label{Eq:R}
\end{equation}
where $\it{\Gamma}$ is the wave power, $t_{\rm{dur}}$  is the time duration of the wave train, and subscripts ``$\rm{i}$'' and ``$\rm{r}$'' refer to incident and reflected wave, respectively. The quantity $t_{\rm{dur, i}}$ consist of the initial ramp-up period and  the first full cycle of the incident wave, while $t_{\rm{dur, r}}$ is the  period of first cycle of the reflected wave. The definition of $t_{\rm{dur, i/r}}$ is motivated by simulation results given in Fig.~\ref{fig:simulation}d where a single cycle of the reflected wave is produced by an incident wave comprising of an initial ramp-up followed by a full cycle.

\subsubsection{Measurement of coefficient of reflection at $z_{2}$}

\begin{figure}
    \centering
    \includegraphics[scale=0.75]{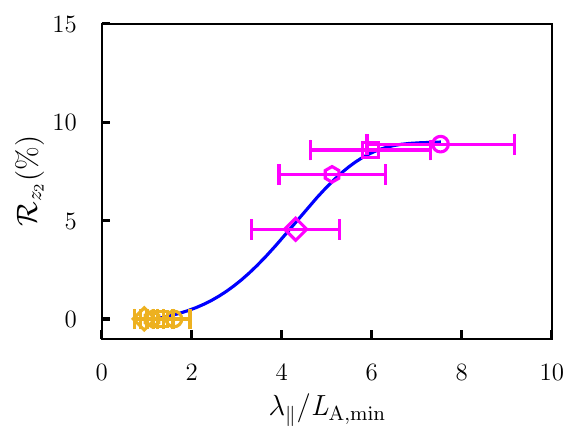}
    \caption{The variation of the coefficient of reflection of Alfv\'en wave measured at $z_{2}$  vs.\ $\lambda_{\parallel}/L_{\rm{A,min}}$. The vertical error bars are smaller than the symbol size. The value of $\lambda_{\parallel}$ was changed by altering the wave frequency. The different wave frequencies are represented by various symbols, circle = $61~\rm{kHz}$, square = $76~\rm{kHz}$, hexagon = $89~\rm{kHz}$, and diamond = $104~\rm{kHz}$. $L_{\rm{A, min}}$ was varied by changing the $B_{0}$ gradient. The magenta and yellow color corresponds to gradients I and III, respectively. The continuous blue line is the most probable function that connect the data points as per Gaussian process regression.   }
    \label{fig:coeff_ref_z2}
\end{figure}

The values of $\mathcal{R}$ calculated by taking the ratio of the reflected to incident wave power at $z_{2}$ is given by color filled symbols in Fig.~\ref{fig:coeff_ref_z2}, where the magenta and yellow colored data points represent measurements for gradients I and III, respectively. $\mathcal{R}$ is presented as a percentage by multiplying Eq.~\ref{Eq:R} by 100. For convenience we have referred to $\mathcal{R}$ measured at $z_{2}$ as $\mathcal{R}_{z_{2}}$.   

$\mathcal{R}_{z_{2}}$ data points for gradient I shows that longer wavelength waves are reflected more strongly from a $v_{\rm{A}}$ gradient than shorter wavelengths. The wavelengths were changed by varying the wave frequency. For the longest wavelength, $\lambda_{\parallel}= 5.6\pm 1.2 ~\rm{m}$ ($\lambda_{\parallel}/L_{\rm{A, min}} = 7.5 \pm 1.6$), and $\mathcal{R}_{z_{2}}$ is $\approx 9\%$. For the shortest wavelength, $\lambda_{\parallel} = 3.2 \pm 0.7~\rm{m} $ ($\lambda_{\parallel}/L_{\rm{A, min}} = 4.3\pm 1~\rm{m}$), and $\mathcal{R}_{z_{2}}$ is $4.6\%$. Note, the errorbar of $\lambda_{\parallel}$ is due to uncertainty in frequency measurement which in turn is due to the finite temporal length of the incident wave. We estimated the frequency uncertainty using the full width at half maximum of the frequency peak in the Fourier spectrum of the incident wave.   

A comparison of $\mathcal{R}_{z_{2}}$ data points for gradients I and III suggests that the  length scale of the $v_{\rm{A}}$ gradient affects wave reflection.  $\mathcal{R}_{z_{2}}$ is zero for gradient III, as the measurement of wave magnetic field along the $x$ axis spanning the width of the plasma column at $y=0$ did not reveal any signature of a reflected wave from the $v_{\rm{A}}$ gradient. However, $\mathcal{R}_{z_{2}}$ is non-zero for I. The key difference between I and III are their values of $L_{\rm{A, min}}$ because $\lambda_{\parallel}$ for both gradients lie in the same range. $L_{\rm{A, min}}$ for I and III are 0.74~m and 3.24~m, respectively. Note that $L_{\rm{A, min}}$ signifies the degree of inhomogeneity of a $v_{\rm{A}}$ gradient and a small $L_{\rm{A, min}}$ indicates a strong gradient. 

The blue curve in Fig.~\ref{fig:coeff_ref_z2} showing the variation of $\mathcal{R}_{z_{2}}$ vs.\ $\lambda_{\parallel}/L_{\rm{A, min}}$ as predicted by Gaussian process regression (GPR), a class of machine learning algorithms \citep{Rasmussen2005}. GPR is a Baysean non-parametric regression technique that predicts a probability distribution over possible functions that fit a set of discrete data points. The  blue curve in Fig.~\ref{fig:coeff_ref_z2} gives mean of the probability distribution i.e., most probable characterization of the data \citep{scikit-learn, bose2022}.

\subsubsection{Estimating coefficient of reflection in the gradient}

\begin{figure}
    \centering
    \includegraphics[scale=0.75]{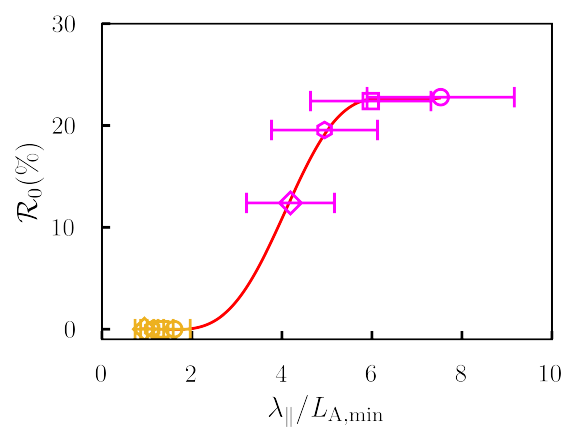}
    \caption{The variation of the coefficient of reflection of Alfv\'en wave estimated at the site of the reflector in the gradient, $\mathcal{R}_{0}$, vs.\ $\lambda_{\parallel}/L_{\rm{A,min}}$. The vertical error bars are smaller than the symbol size. The continuous red line is the most probable function that connect the data points as per Gaussian process regression.  }
    \label{fig:coeff_ref}
\end{figure}

The values of $\mathcal{R}_{z_{2}}$ in Fig.~\ref{fig:coeff_ref_z2} are the lower limit of the coefficient of reflection. This is because the incident Alfv\'en wave damps as it propagates from $z_{2}$ to the location of the reflector in the gradient, causing the incident wave energy to be smaller at the location of the reflector than at $z_{2}$.  Similarly, the reflected wave damps as it propagates from the site of the reflector in the gradient, causing the reflected wave energy to be smaller at $z_{2}$.   As a result, the incident wave energy is overestimated while the reflected wave energy is underestimated at $z_2$ compared to the corresponding energies at the site of the reflector in the gradient.

We have used a simple model to estimate the incident and reflected wave energy at the site of the wave reflector to make a zeroth order correction to the directly measured values of $\mathcal{R}_{z_{2}}$.  In the model, the incident wave energy at $z_{2}$ ($\mathcal{E}_{\rm{i}}$) is related to the incident wave energy at the site of wave reflector ($\mathcal{E}_{\rm{i, 0}} $) by 
\begin{equation}
    \mathcal{E}_{\rm{i, 0}}\approx \mathcal{E}_{\rm{i}}\exp{\left( -2\Delta z/d \right)},
\label{Eq:dampint_rate}
\end{equation}
where $\Delta z$ is the distance between $z_{2}$ and location of the reflector, and $d$ is the collisional damping length of Alfv\'en wave, i.e., the length over which the Alfv\'en wave amplitude decreases to $1/e$ of its initial value. Note, the factor 2 appears in the exponential term  of Eq.~(\ref{Eq:dampint_rate}) because $\mathcal{E} \propto b^2$ and the damped amplitude of $b \propto \exp{\left( -\Delta z/d \right)} $.  Similarly, the reflected wave energy at the reflector ($\mathcal{E}_{\rm{r, 0}}$) and at $z_{2}$ ($\mathcal{E}_{\rm{r}}$) are related by   
\begin{equation}
    \mathcal{E}_{\rm{r}} \approx \mathcal{E}_{\rm{r,0}}\exp{\left( -2\Delta z/d \right)}. 
\end{equation}
Therefore, the coefficient of reflection at the site of the reflector ($\mathcal{R}_{0}$) is given by 
\begin{equation}
    \mathcal{R}_{0}\approx \frac{\mathcal{E}_{\rm{r}, 0}/t_{\rm{dur,r}}}{\mathcal{E}_{\rm{i}, 0}/t_{\rm{dur,i}}}\approx \mathcal{R}_{z_{2}}\exp{\left(  4 \Delta z/d\right)}.
\end{equation}

The collisional damping length is estimated using  the two-fluid dispersion relation of Alfv\'en wave for uniform plasma  in the $\bar{\beta} \gg 1$ limit, \citep{vranjes2006unstable, gigliotti2009generation, bose2019measured} 
\begin{equation}
\omega^{2} - k_{\parallel}^2 v_{\rm A}^2 \left(1-\bar{\omega}^2 + k_{\perp}^2 \rho_{\rm s}^2 \right)+i \omega k_{\perp}^2 \delta_{e}^2 \nu_{e} = 0. 
\label{Eq:collisional_KAW}
\end{equation}  
\noindent
This quadratic equation can be solved algebraically, and the imaginary root, $k_{\rm Im, \parallel}$, gives the damping length, 
\noindent
\begin{align}
d& = \frac{1}{k_{\rm Im, \parallel}}   \notag \\ 
& = \frac{\sqrt{2} v_{\rm A}}{\omega} \left[ \frac{\sqrt{1+k_{\perp}^4 \delta_{e}^4 \left( \nu_e /\omega \right)^2}-1}{1-\bar{\omega}^2 + k_{\perp}^2 \rho_{\rm s}^2} \right]^{-1/2}. 
\label{Eq:ki_collisional_KAW}
\end{align}  
\noindent
The dominant perpendicular wavelength of the incident wave, $\lambda_{\perp}=21.78~\rm{cm}$. The average value of the plasma parameters between $z_{2}$ and the reflector are employed for estimating $d$ using Eq.~\ref{Eq:ki_collisional_KAW}. The site of reflector is taken to be location of the strongest part of the $v_{\rm A}$ gradient, which is at $z=6.2~\rm{m}$, therefore, $\Delta z = 2.7~\rm{m}$. The average $B_{0}$ between $z_{2}$ and $z=6.2~\rm{m}$ is 623~G. The values of $d$ for 61, 76, 89 and 104~kHz waves are 11.4, 11.2, 11, and 10.6~m, respectively.

The resulting value of $\mathcal{R}_{0}$ has a monotonic variation with respect to $\lambda_{\parallel}/L_{\rm{A, min}}$ as shown in Fig.~\ref{fig:coeff_ref}. The data points are shown by symbols while the red curve is most probable characterization of the data points as per GPR. The $\mathcal{R}_{0}$ vs.\ $\lambda_{\parallel}/L_{\rm{A, min}}$ curve shows that as the  inhomogeneity increases, so does the reflection of Alfv\'en waves. 

The values of $\mathcal{R}_{0}$ are approximate quantities as we estimated $d$ of the Alfv\'en wave in the gradient using the dispersion relation for a uniform plasma. A comprehensive analysis to accurately determine $d$   in a strong non-WKB gradient relevant to our experiments is beyond the scope of this paper. However, past experiments of \citet{bose2019measured} does provide an insight on the damping of Alfv\'en waves in a strong non-WKB gradient. They found the energy reduction of an Alfv\'en wave in a strong non-WKB gradient to be more than in an almost uniform plasma suggesting that the damping length is smaller (i.e. damping is more pronounced) in the gradient than in the uniform case.     This implies that for the uniform plasma the values of $d$ used to calculate $\mathcal{R}_{0}$  overestimates the damping length in the gradient and thus underestimates $\mathcal{R}_{0}$. Therefore, $\mathcal{R}_{0}$ in Fig.~\ref{fig:coeff_ref} is likely to be a lower bound on the  coefficient of reflection after correcting for collisional damping.

\begin{figure*}
    \centering
    \includegraphics[scale=0.89]{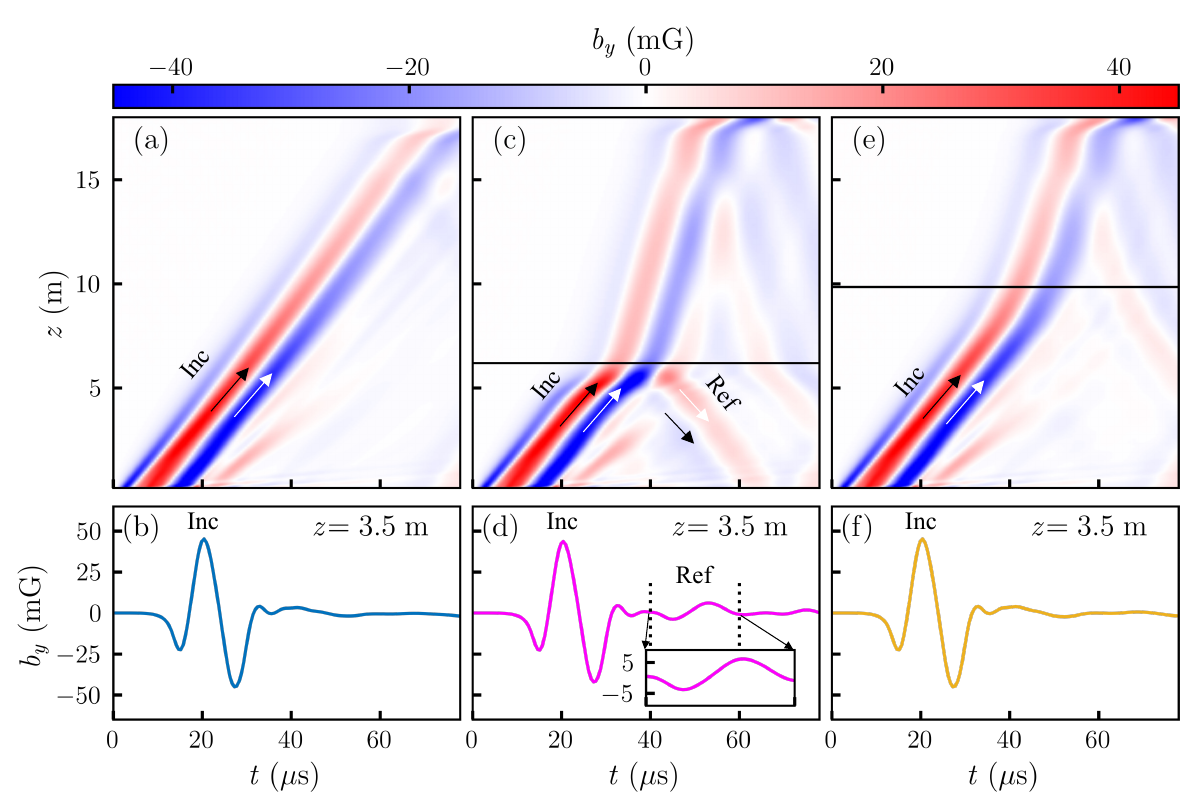}
    \caption{(Top row) 2-D color plots of $b_y(x=y=0)$ displaying the propagation of the incident (Inc) Alfv\'en wave and its reflection (Ref) in the $tz$-plane for gradient (a) U, (c) I, and (e) III. The $z$ axis extend from 0.2 to 18~m, while the $t$ axis ranges from 0 to 78~$\mu\rm{s}$. The deep red color  implies that the oscillating $b_{y}$ is at its maximum positive value of the wave peak, while the deep blue indicates $b_{y}$ is at the minimum of the wave trough. The arrows trace the phase of the wave. The black horizontal line in (c) and (e) show the $z$ location of $L_{\rm{A,min}}$ in the $v_{\rm{A}}$ gradient. (Bottom row) The time variation of $b_{y}$ obtained by taking a horizontal cut at $z=3.5\:\rm{m}$ for gradients (b) U, (d) I, and (f) III.  The inset figure in (d) show $b_{y}$ vs.\ $t$ between 40 to 60~$\mu \rm{s}$.}
    \label{fig:simulation}
\end{figure*}

\section{Simulations}\label{sec:two_fluid}

\subsection{Overview}

We employ the five-moment, two-fluid model \citep{Hakim:2006,Wang:2015,Wang:2020} within the {\tt Gkeyll} simulation framework to model Alfv\'{e}n-wave reflection in LAPD. The five-moment model evolves equations for density, momentum, and isotropic pressure for each species, which are coupled through Maxwell's equations, and includes the displacement current. We model the full LAPD domain in Cartesian coordinates with reflecting boundary conditions in each dimension, $-0.6~\rm{m}$~$ \le~x,~y~\le~0.6 ~\rm{m}$ and $-10~ \rm{m}$~$ \le~z~\le~18~ \rm{m}$, and employ $n_x = n_y = 64$ and $n_z = 700$ cells. We initialize a Helium plasma, $m_i = 4m_p$, with $m_p = 1.67 \times 10^{-24}$ g and a reduced mass ratio, $m_i / m_e = 100$, uniform temperatures and densities based on fiducial values from the experiments, $T_e = 7$~eV, $T_e/T_i = 5$, and $n=n_e = n_i = 7\times10^{12}~ \rm{cm}^{-3}$. The speed of light for simulations is taken to be  $c \simeq 9.5\times10^6~\rm{m\,s^{-1}}$. Three background, axial magnetic field profiles are employed corresponding to field profiles U, I,  and III in Fig.~\ref{fig:exp_set_up}. We drive the plasma with a model of the Arbitrary Spatial Waveform (ASW) antenna placed at $z = 0~\rm{m}$ \citep{thuecks2009tests,Drake:2013}. The antenna parameters are $J = 10~\rm{kA\,m^{-3}}$, $f_{\rm{ant}} = 76.4~\rm{kHz}$, and $\lambda_{\perp \rm ant} = 21.78~\rm{cm}$, which drives an Alfv\'{e}n wave with $\lambda_\parallel \simeq 4.4~\rm{m}$. The antenna is driven for a time $t_{\rm{drive}} = 1.5/f_{\rm{ant}}$. This includes ramping up and down periods, each of length $t_{\rm ramp} = 0.25/f_{\rm ant}$. We note that we both extend the LAPD-model domain to $z = -10~\rm{m}$ and flatten the background magnetic field gradient in this region to reduce spurious reflections from behind the antenna.

\subsection{Results}

\subsubsection{Alfv\'en wave through uniform $B_{0}$}

Fig.~\ref{fig:simulation}(a) shows the wave propagation along the cylindrical axis of LAPD in the $tz$-plane for gradient U. The color gives the $y$-component of the wave magnetic field, which in turn also indicates the phase of the wave.  The slope of the band traced by the phase of the wave indicates the propagation direction of the wave; positive and negative slope corresponds to $+z$ and $-z$ direction, respectively.  In Fig.~\ref{fig:simulation}(a), the slope of the band traced by the phase of the wave  is positive,  consistent for an incident wave propagating away from the antenna. Fig.~\ref{fig:simulation}(b) shows the time variation of the incident wave at $z=3.5~\rm{m}$. Unlike the experiment, we do not see a wave reflected from the far end in the time series data because the simulation was carried out up to $\approx 78\;\mu s$, and the wave reflected from the far end requires a longer time to reach $z=3.5~\rm{m}$.

\subsubsection{Alfv\'en wave reflection from a strong $v_{\rm{A}}$ gradient}
Fig.~\ref{fig:simulation}(c) shows the wave propagation in the $tz$-plane for gradient I, where we observe experimentally wave reflection from the $v_{\rm{A}}$ gradient. Initially, from $t=0$ to $\lesssim 40~\rm{\mu s}$ and $z \lesssim  5~\rm{m}$, the wave propagates in $+z$ direction as expected for an incident wave. Upon reaching the gradient region, where $L_{\rm{A}}\approx L_{\rm{A, min}}$, a portion of the incident wave is reflected. The negatively sloped arrows trace the phases of the wave reflected from the $v_{\rm{A}}$ gradient. 

We observe that reflection from a $v_{\rm{A}}$ gradient introduces a $180^{\circ}$ phase difference between the wave magnetic field of the incident and the reflected wave. In Fig.~\ref{fig:simulation}(c), the deep blue band of the incident wave, marked by the positively sloped white arrow, correlates to the negatively sloped red band (reflected wave), marked by the negatively sloped white arrow.  Similarly, the positively sloped deep red band of the incident wave corresponds to the negatively sloped blue band of the reflected wave, as shown by $ \rm{follow\;the \;black \;arrows\; in\; Fig.~\ref{fig:simulation}(c)} $. The change in the color between the corresponding phases of the incident and the reflected wave represents a $180^{\circ}$ phase difference. We performed a second test to check the phase difference. Simulations using an incident wave having only a strong negative $b_{y}$ (i.e. no strong positive $b_{y}$) gave rise to a reflected wave with only a positive $b_{y}$, further demonstrating a $180^{\circ}$ phase difference between incident and reflected wave due to reflection. See Appendix~\ref{appen_B} for  the additional details on this second test.

The time variation of $b_{y}$ before the gradient at $z=3.5~\rm{m}$ in Figs.~\ref{fig:simulation} (c) and (d) show that the first low amplitude trough of the incident wave does not produce a strong detectable reflected wave. The reasons for this observation may be related to the process associated with Alfv\'en wave reflection at the boundary, such as establishing currents at the boundary to support a reflected Alfv\'en wave. A discussion on the  interaction of the Alfv\'en wave with the $v_{\rm{A}}$ gradient boundary is beyond the scope of this paper and will be discussed elsewhere.

\subsubsection{Alfv\'en wave incident on a weak $v_{\rm{A}}$ gradient}
Fig.~\ref{fig:simulation}(e) shows the wave propagation in the $tz$-plane for case III, i.e., the $B_{0}$ profile with gradient III. We do not observe any wave reflection from the $v_{\rm{A}}$ gradient, but the transmitted wave is reflected from the far end of the simulation domain. The $b_{y}$ time series data in Fig.~\ref{fig:simulation}(f) only shows a well-formed incident wave and does not exhibit any signature of a reflected wave from the $v_{\rm{A}}$ gradient.

\section{Discussion}\label{sec:discussion}

\begin{figure}
    \centering
    \includegraphics[scale=0.87]{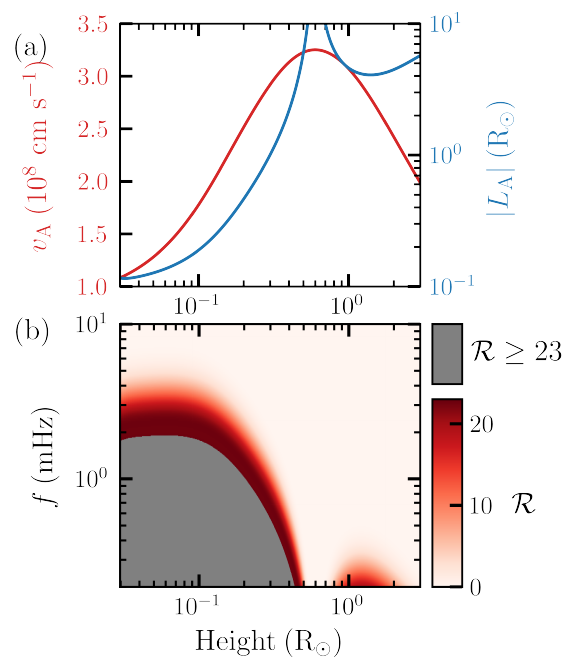}
    \caption{(a) The variation of Alfv\'en speed, $v_{\rm{A}}$,  and length scale of the $v_{\rm{A}}$ gradient in a coronal hole vs.\ height from the surface of the Sun. (b) A  colormap showing the estimated coefficient of reflection, $\mathcal{R} (\%)$, for different wave frequencies ($0.2 - 10$~mHz) at different heights above the surface of the Sun in a coronal hole.  Note that Alfv\'en waves are reflected strongly at low heights ($\lesssim 0.13~\rm{R_{\odot}}$) where $v_{\rm{A}}$ gradient is strong (see main text).  The words ``strong reflection" in this context implies wave reflection that may be enough to generate sufficient heating of the plasma via the wave turbulence mechanism.        }
    \label{fig:va_sun}
\end{figure}

Our experimental results have implication on the heating of coronal holes. A significant temperature increase of coronal hole plasma occurs at low heights \citep{ko1997empirical,landi2008off}, a region where $v_{\rm{A}}$ is highly non-uniform. The spatial variation of $v_{\rm{A}}$ is shown by the red curve in Fig.~\ref{fig:va_sun}(a).  We used approximate fits of density and magnetic field given by \citet{cranmer2005generation} to calculate $v_{\rm{A}}$.

A comparison of the wave inhomogeneity parameter ($\lambda_{\parallel} / L_{\rm{A}}$) for a coronal hole with that of the laboratory suggest that $\mathcal{R}$ can be greater than $23\%$ at low heights. The spatial variation of $v_{\rm{A}}$ and $L_{\rm{A}}$ in a coronal hole are given by the red and  blue curves, respectively, in Fig.~\ref{fig:va_sun}(a). The   $\lambda_{\parallel}$ at different heights in a coronal hole for various wave frequencies $f$, can be  estimated using $\lambda_{\parallel}=v_{\rm{A}}/f$. The values of $f$ vary from 0.2 to 10~mHz \citep{morton2015investigating}. The resulting values of $\lambda_{\parallel}/L_{\rm{A}}$ were used to obtain estimates of $\mathcal{R}$ for a coronal hole by comparing $\lambda_{\parallel}/L_{\rm{A}}$ of a coronal hole with $\lambda_{\parallel}/L_{\rm{A, min}}$ of the laboratory (red curve in Fig.~\ref{fig:coeff_ref}). Fig.~\ref{fig:va_sun}(b) shows $\mathcal{R}$ for different values of $f$ at various heights above the surface of the Sun. The different shades of red show the values of $\mathcal{R}$ from 0 to $23\%$, while the grey indicates $\mathcal{R} \geq 23\%$. The reason for using the grey color is that in a portion of the frequency-height plane the coronal hole values of $\lambda_{\parallel}/L_{\rm{A}}$ are greater than the maximum value of $\lambda_{\parallel}/L_{\rm{A, min}}=7.5$ attained in the laboratory. Since, the coefficient of reflection has a monotonic dependence on $\lambda_{\parallel}/L_{\rm{A, min}}$ in the laboratory with a maximum value of $23\%$ we assumed that $\mathcal{R}$ is $\geq 23\%$ in a coronal hole for  $\lambda_{\parallel}/L_{\rm{A}}>7.5$.   

A comparison of the data in Fig.~\ref{fig:va_sun}b with solar observations and solar  simulations suggests that wave reflection from a smooth $v_{\rm{A}}$ gradient may be sufficient to significantly heat the plasma at low heights up to $0.13~\rm{R_{\odot}}$ in coronal holes via the wave turbulence model. Solar observation suggest that waves have sufficient energy to heat a coronal hole \citep{mcintosh2011alfvenic}. Furthermore, measurement of the wave frequency spectrum by \citet{morton2015investigating}  indicates that bulk of the wave energy occur at frequencies below $\sim 3~\rm{mHz}$ in a coronal hole. Simulations of \citet{asgari2021effects}  suggests that a reflection coefficient of $\sim10\%$  may be sufficient to heat the plasma at heights below $3~\rm{R_{\odot}}$ in coronal holes. They invoked density fluctuations to reflect $\sim 10\%$ of the wave power. However, our laboratory results suggest that a smooth $v_{\rm A}$ gradient relevant to low heights in coronal holes can reflect more than $10\%$ of wave power. In Fig.~\ref{fig:va_sun}(b) $\mathcal{R}$ is greater than $10\%$ for $f\leq 3~\rm{mHz}$ for a height up to $0.13~\rm{R_{\odot}}$. Therefore, wave turbulence driven by nonlinear interaction between outward propagating waves from the Sun's surface and inward reflected waves may heat a coronal hole plasma enough to sustain $10^{6}~\rm{K}$ temperature in lower corona up to a height of $0.13~\rm{R_{\odot}}$.

Although $\mathcal{R}$ tends to decrease above $0.13~\rm{R_{\odot}}$, reflection may still continue to play a role in heating the plasma at heights up to  $\approx 0.5~\rm{R_{\odot}}$. The frequency spectrum of Alfv\'enic waves measured by \citet{morton2015investigating} suggest  the wave energy is concentrated in  frequencies at the lower end of the spectrum. In their measurement, the wave energy peaked at $\sim 0.2~\rm{mHz}$, which was the lowest frequency in their dataset. In Fig.~\ref{fig:va_sun}(b) waves of $0.2~\rm{mHz}$ frequency have $\mathcal{R} \geq 10\%$ up to a height of $\approx 0.5~\rm{R_{\odot}}$. The value of $\mathcal{R}$ for wave frequencies above 0.2~mHz progressively decreases from $0.13~\rm{R_{\odot}}$ to $\approx 0.5~\rm{R_{\odot}}$. These two features of Fig.~\ref{fig:va_sun}(b) suggest that wave reflection from the smooth $v_{\rm{A}}$ gradient will play a partial role in heating the plasma via a wave turbulence mechanism but that the the role of reflection is expected to progressively decrease with height from  $0.13~\rm{R_{\odot}}$ to $\approx 0.5~\rm{R_{\odot}}$. At heights above $\approx 0.5~\rm{R_{\odot}}$, reflection from the smooth $v_{\rm{A}}$ gradient is rather weak and the reflected wave may not be of consequence in heating the plasma.

\section{ Summary}\label{sec:summary}

In this paper, we report the first experimental detection of a reflected Alfv\'en wave from a strong $v_{\rm{A}}$ gradient using a series of experiments in solar-like plasmas. Our results show that a strong $v_{\rm{A}}$ gradient is necessary to reflect Alfv\'en waves, and that the reflected wave energy  increases for incident waves having longer wavelengths. The trend in the laboratory wave reflection data agree with the solar observation of \citet{morton2015investigating}, where they showed that the ratio of the outward propagating wave power to the inward propagating wave power increases for longer wavelength waves, i.e., smaller wave frequencies. This indicates that the counter-propagating waves observed by \citet{morton2015investigating} may be due to wave reflection, supporting their hypothesis.

Two-fluid simulations using the {\tt Gkeyll} code qualitatively agree with and support the experimental detection of a reflected Alfv\'en wave. In the simulations, we excited Alfv\'en waves using a model of the ASW antenna. We traced the propagation of Alfv\'en waves along the plasma column through a uniform $B_{0}$ profile, as well as strong and weak $v_{\rm{A}}$ gradients.  Like the experiments, the simulations show that a strong $v_{\rm{A}}$ gradient reflects, while a weak $v_{\rm{A}}$ gradient does not. In the future, we will make quantitative comparison after developing a model of an orthogonal ring antenna for simulations similar to the one used in experiment.

We experimentally measured $\mathcal{R}$ for different values of $\lambda_{\parallel}/L_{\rm{A, min}}$ and presented two sets of values of $\mathcal{R}$, i.e., $\mathcal{R}_{z_{2}}$ and $\mathcal{R}_{0}$. $\mathcal{R}_{z_{2}}$ is determined by dividing the measured reflected wave power just after the wave exits from the gradient region by the incident wave power  that enters the gradient region. The gradient region is of finite spatial extent, and since simulations showed that wave reflects from the vicinity of the strongest part of the gradient, we estimated the coefficient of reflection near the strongest part, $\mathcal{R}_{0}$, using $\mathcal{R}_{z_{2}}$ and a model to take into account collisional damping between the edge of the gradient region and the center of the gradient region. The values of $\mathcal{R}_{z_{2}}$ and $\mathcal{R}_{0}$ were found to increase with increasing $\lambda_{\parallel}/L_{\rm{A, min}}$.

A comparison of the laboratory results and solar parameters suggest that Alfv\'en waves will be reflected by the smooth $v_{\rm{A}}$ gradient at heights up to $0.5~\rm{R_{\odot}}$ in coronal holes. Furthermore, solar models suggest that the coefficient of reflection may be sufficient to generate enough inward propagating waves at low heights in a coronal hole ( $\lesssim 0.13~\rm{R_{\odot}}$)  to turbulently heat the plasma at these heights. The role of wave reflection from the smooth $v_{\rm{A}}$ gradient in heating the plasma is expected to progressively diminish from a height of $\approx 0.13$  to $0.5~\rm{R_{\odot}}$. In addition, at heights of $\approx 0.5~\rm{R_{\odot}}$ or above, reflection from this gradient may not be adequate to generate enough plasma heating via wave turbulence mechanism requiring other processes like density fluctuation to enhance wave reflection \citep{shoda2019three, asgari2021effects}.

\appendix

\section{Phase relationship between incident and reflected wave}\label{appen_B}

\begin{figure}
    \centering
    \includegraphics[scale=0.8]{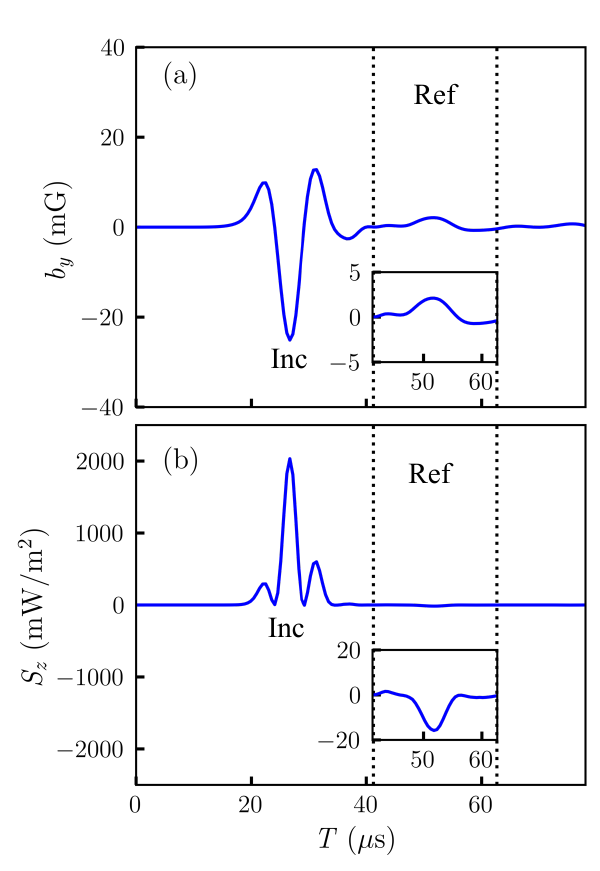}
    \caption{ Time variation of (a) $b_{y}$ and (b) $S_{z}$ for profile I at $x=y=0$ and $z=3.5~\rm{m}$ showing the incident wave referred to as ``Inc'' and the reflected wave denoted by ``Ref''. }
    \label{fig:sim_PT}
\end{figure}

We performed additional simulations to  test the phase relationship between the incident and reflected waves. Profile I was used for the simulations, where the Alfv\'en wave propagated from a region of low to high velocity. An incident wave having a strong negative $b_{y}$ was found to produce a reflected wave having a positive $b_{y}$, as shown in Fig.~\ref{fig:sim_PT}(a). The direction of propagation of the wave was confirmed by calculating the $z$ component of the Poynting vector, $S$, given in Fig.~\ref{fig:sim_PT}(b). A positive value of $S_{z}$ signifies propagation in the $+z$ direction, which is consistent for an incident wave. A reflected wave propagating in the $-z$ direction has a negative  $S_{z}$ value.  Therefore, Fig.~\ref{fig:sim_PT}(a) and (b) demonstrates that Alfv\'en wave reflection from gradient I introduces a $180^{\circ}$ phase difference between the incident and reflected wave. 

The result of the Alfv\'en wave reflection simulations agree with the predictions of theory of reflection of light \citep{griffiths2015introduction}. According to which, the phase of the magnetic field of a light wave incident on a gradient from the low velocity side is inverted by $180^{\circ}$ upon reflection. The reason for the similarity between Alfv\'en and light waves may be because both are electromagnetic waves.  

\vspace{12pt}

\section*{Acknowledgements}
This work was supported by US DOE Frontier Plasma Science program and General plasma science program funded by contract number DE- AC0209CH11466. J.M.T. was additionally supported by the NSF CSSI (Cyberinfrastructure for Sustained Scientific Innovation) program, grant number 2209471. D.W.S. and M.H. were supported by the DOE grant DE-SC0021261. J. Juno was supported by the U.S. Department of Energy under Contract No. DE-AC02-09CH1146 via an LDRD grant. The experiments were performed at the Basic Plasma Science Facility (BaPSF), which is a colaborative research facility supported by the DOE and NSF, with major facility instrumentation developed via an NSF award AGS–9724366. The simulations presented in this article were performed on computational resources managed and supported by Princeton Research Computing, a consortium of groups including the Princeton Institute for Computational Science and Engineering (PICSciE) and the Office of Information Technology's High Performance Computing Center and Visualization Laboratory at Princeton University.

The United States Government retains a non-exclusive, paid-up, irrevocable, world-wide license to publish or reproduce the published form of this manuscript, or allow others to do so, for United States Government purposes.

\section*{Data availability statement}
The experimental data presented in this paper will be made available upon publication using Princeton's data repository. {\tt Gkeyll} is open source and can be installed by following the instructions on the {\tt Gkeyll} website (http://gkeyll.readthedocs.io). The input and LAPD gradient files for the {\tt Gkeyll} simulations presented here are available in the following GitHub repository, https://github.com/ammarhakim/gkyl-paper-inp.

\bibliography{sample631}{}
\bibliographystyle{aasjournal}


\end{document}